# Inverse variational problem for non-standard Lagrangians

Aparna Saha[1*] and B Talukdar[2]

[1] Department of Physics, Visva-Bharati University, Santiniketan 731235, India

[2] Nutan Palli, Bolpur 731204, India

[*] Email address: aparna_phyvb@yahoo.co.in

**Abstract.** The non-standard Lagrangians (NSLs) for dissipative-like dynamical systems were introduced in an ad hoc fashion rather than being derived from the solution of the inverse problem of variational calculus. We begin with the first integral of the equation of motion and solve the associated inverse problem to obtain some of the existing results for NSLs. In addition, we provide a number of alternative Lagrangian representations. The case studies envisaged by us include (i) the usual modified Emden-type equation, (ii) Emden-type equation with dissipative term quadratic in velocity and (iii) Lokta-Volterra model. We point out that our method is quite general for applications to other physical systems.

**1. Introduction**

In the calculus of variations one deals with two types of problems, namely, the direct and inverse problems of mechanics. The direct problem is essentially the conventional one in which one assigns a Lagrangian to the physical system and then computes the equations of motion using the Euler-Lagrange equations. On the contrary, the inverse problem begins with the equations of motion and then constructs the Lagrangian of the system by a strict mathematical procedure [1].

The Lagrangian $L$ of an autonomous differential equation is expressed as $L = T - V$ where $T$ is the kinetic energy of the system modeled by the equation and $V$, the corresponding potential function. In recent years, a new type of Lagrangian functions have been proposed for dissipative-like autonomous differential equations [2]. These do involve neither $T$ nor $V$, but yet yield the equations of motion via Euler-Lagrange equations. Consequently, these Lagrangians were qualified as 'non-standard'. It appears that the non-standard Lagrangians do not have a natural space in the theory of inverse variational problem. In other words, the basic question whether the proposed expressions for NSLs can be obtained from the differential equations which they represent has remained largely unanswered. In this work we shall deal with this problem and present expressions of NSLs for a number of equations which are of considerable current interest.

The inverse problem in classical mechanics was solved by Helmholtz towards the end of the nineteenth century [3]. The use of Helmholtz machinery to solve an inverse problem requires the knowledge of algebro-geometric theories [4]. Recently, Nucci and Leach [5] made use of a method devised by Jacobi to construct Lagrangians for a wide class of second-order differential equations. Their method consists in finding a Jacobi Last Multiplier the concept of which was introduced by Jacobi long before the work of Helmholtz [6]. In this context we identify a solution of the inverse variational problem due to Lopez [7] who provided a method to construct the Lagrangian function for an N dimensional second-order autonomous differential equation from its first integral. Here the Lagrangian is given by

$$L(\vec{x}, \vec{v}) = \frac{1}{N} \sum_{i=1}^{N} v^i \int^{v^i} \frac{C^i(\vec{x}, \xi) d\xi}{\xi^2} \qquad (1)$$

with the coordinate $\vec{x}$ and velocity $\vec{v}$ written as

$$\vec{x} = \{x^1, x^2, \ldots x^N\} \quad \text{and} \quad \vec{v} = \frac{d\vec{x}}{dt} = \{v^1, v^2, \ldots v^N\} \,. \qquad (2)$$

In (1) each of the $C^i$'s stands for the appropriate first integral of the $N$-dimensional equation. The expression of the Lagrangian in (1) was obtained by deriving a first-order partial differential equations for the constant of the motion and ultimately relating it to the corresponding Jacobi integral [8]. In treating the inverse problem that leads to non-standard Lagrangians we shall make judicious use of the result given in (1).

The modified Emden-type equation is given by [9]
$$\ddot{x} + \alpha x \dot{x} + \beta x^3 = 0, \; x = x(t) \tag{3}$$
with $\alpha$ and $\beta$ as arbitrary constants This equation with $0 < \alpha < 1$ plays a role in several applicative contexts [10] Two special equations as obtained from (3), namely,
$$\ddot{x} + 3kx\dot{x} + k^2 x^3 = 0, \; \beta = \alpha^2/9 \tag{4}$$
and
$$\ddot{x} + 4kx\dot{x} + 2k^2 x^3 = 0, \; \beta = \alpha^2/8 \tag{5}$$
have been found [11 – 13] to follow from the non-standard Lagrangians
$$L = \frac{1}{\dot{x} + kx^2} \tag{6}$$
and
$$L = \ln(\dot{x} + kx^2) \tag{7}$$

In section 2 we shall derive the results in (6) and (7) from the solution of the inverse problem for (4) and (5). In addition we shall present a number of inequvalent or alternative Lagrangians for these equations. The alternative Lagrangians are not connected by gauge terms, yet they give the same equation of motion [14]. The existence of alternative Lagrangians have deep consequences in physical theories. For example, there can arise ambiguities in the association of symmetries with conservation laws. Moreover, the same physical system, via alternative Lagrangian descriptions, can lead to entirely different quantum mechanical systems.

Recently, we demonstrated that a modified Emden-type equation written as [15]
$$\ddot{x} + \alpha x \dot{x}^2 + \beta x^3 = 0 \tag{8}$$
is solvable for all values of the parameters $\alpha$ and $\beta$. Here the dissipative term is quadratic in velocity. In section 3 we show that (8) follows from both standard and non-standard type Lagrangians.

Works on the dynamics ecological systems by the use of canonical procedure were begun in mid 1970s [16]. Nucci and Tamizhmani [17] made use of the approach developed in ref.5 to provide expressions for Lagrangians functions for a number of biological models which may be of interest for studying population dynamics by the canonical procedure. We devote section 4 to examine if the approach followed by us can be used to obtain Lagrangians for biological models. To that end we deal with the Lotka-Volterra model [16] which is often used to simulate interactions between two or more populations. The primary results obtained by us are of the non-standard type as found in ref. 16. However, the constructed alternative Lagrangians belong to the standard type, Finally in section 5 we make some concluding remarks.

## 2. Modified Emden-type Equations (MEEs)

Equation (4) is equivalent to two first-order differential equations given by
$$\dot{x} = y, \; y = y(t) \tag{9}$$
and
$$\dot{y} = -(3kxy + k^2 x^3) \; . \tag{10}$$
From (9) and (10) we write
$$y' = -\frac{3kxy + kx^3}{y}, \; y' = \frac{dy}{dx} \; . \tag{11}$$
Fortunately, the solution of (11) can be obtained in the analytic form to get
$$\dot{x} = \frac{1}{3}k(e^{\frac{2c}{9}} - 3kx^2) + e^{\frac{c}{9}}\sqrt{e^{\frac{2c}{9}} - 3kx^2} \tag{12}$$
where $c$ is a constant of integration. By defining a new constant $c_1 = e^{\frac{c}{9}}$, we can solve (12) to write

$$c_1(x,\dot{x}) = -\sqrt{\frac{3}{k}} \frac{\dot{x}+kx^2}{\sqrt{2\dot{x}+kx^2}}. \qquad (13)$$

For our one dimensional system (1) reads

$$L(x,\dot{x}) = \dot{x}\int^{\dot{x}} \frac{C(x,\xi)}{\xi^2} d\xi. \qquad (14)$$

It is straightforward to verify that the choice $C(x,\dot{x}) = \dfrac{1}{c_1^2(x,\dot{x})}$ when substituted in (14) leads to the inverse type of the non-standard Lagrangian (6). Similarly, from (14) we can obtain

$$L = \sqrt{2\dot{x}+kx^2} \qquad (15)$$

and 
$$L = \frac{3}{2k}\dot{x}\ln(2\dot{x}+kx^2) - 3x^2 \qquad (16)$$

for the choices $C(x,\dot{x}) = c_1(x,\dot{x})$ and $C(x,\dot{x}) = c_1^2(x,\dot{x})$. In this way we find many more Lagramgians for the modified Emden-type equation (4). As with the result in (6), the Lagrangians (15) and (16) also belong to the non-standard class. Moreover, these are not related to that in (6) by gauge terms. Thus they stand for nonstandard alternative Lagrangians of the system. In fact, by the procedure followed here, one will be able to construct a family of such inequivalent Lagrangians.

For (5) the constant of the motion similar to that in (13) is given by

$$c(x,\dot{x}) = \ln(\dot{x}+kx^2) + \frac{kx^2}{\dot{x}+kx^2}. \qquad (17)$$

It is interest to note that for given values of $c(x,\dot{x})$, (13) can be solved to write $\dot{x}$ as a function of $x$ while (17) involves $\dot{x}$ and $x$ in an essentiaslly non-algebraic way. This provides an awkward analytical constraint to solve (5) by simple quadrature. This was why Chandrasekar et.al [18] solved (5) by treating it as a Hamiltonian system and then taking recourse to the use of a canonical transformation. Recently, we derived a method to find solution of this equation by judicious modification of an approach used to solve Liénard type differential equations [13,19]. However, there appear no problem to obtain the Lagrangian of (5) by use of (17). For example, identifying $c(x,\dot{x})$ of (17) with $C(x,\dot{x})$ we obtain from (14) the non-standard logarithmic Lagrangian (7). In this case if we choose $C(x,\dot{x}) = c^2(x,\dot{x})$ we arrive at an inequivalent Lagrangian

$$L = \frac{\dot{x}}{\dot{x}+kx^2} + (\ln(\dot{x}+kx^2))^2 + 2\ln(\dot{x}+kx^2). \qquad (18)$$

As in the case of (4) we can also construct a Lagrangian galore for (5).

## 3. MEE with friction quadratic in velocity

For (8) we have found that the constant of the motion as

$$c(x,\dot{x}) = \frac{e^{\alpha x^2}(\alpha^2 \dot{x}^2 + \alpha\beta x^2 - b)}{\alpha^2}. \qquad (19)$$

Using $C(x,\dot{x}) = c(x,\dot{x})$ in (14) we obtain

$$L = e^{\alpha x^2}(\frac{1}{2}\dot{x}^2 - \frac{1}{2}(\frac{\beta}{\alpha}x^2 - \frac{\beta}{\alpha^2})). \qquad (20)$$

If we choose to work with $C(x,\dot{x}) = 1/c(x,\dot{x})$, we shall get

$$L = e^{-\alpha x^2} \left[ \frac{\theta^3 e^{-\alpha x^2} \tan^{-1}\theta}{\dot{x}^2} + \frac{\alpha^2}{\beta(\alpha x^2 - 1)} \right] \quad (21)$$

with $\quad \theta = \dfrac{\alpha \dot{x}}{\sqrt{\beta(\alpha x^2 - 1)}} \quad .\qquad (22)$

Looking at (20) and (21) it is clear that the former represents a standard type Lagangian while (21) belongs to the non-standard class. We have verified that the (8) has only one standard Lagrangian representation and all other alternative Lagrangians are non-standard.

### 4. Lotka-Volterra Model

In ref.16 the Lotka-Volterra model was represented by the set of coupled first-order ordinary differential equations written as

$$\dot{w}_1 = w_1(a + bw_2), \ w_1 = w_1(t) \quad (23)$$

and $\quad \dot{w}_2 = (A + Bw_1) \ . \ w_2 = w_2(t) \ . \quad (24)$

The change of variables

$$w_1 = e^{r_1} \text{ and } w_2 = e^{r_2}, \ r_1 = r_1(t) \text{ and } r_2 = r_2(t) \quad (25)$$

reduces (23) and (24) to

$$\dot{r}_1 = be^{r_2} + a \quad (26)$$

and $\quad \dot{r}_2 = Be^{r_1} + A \quad (27)$

respectively. We now differentiate (26) and (27) to get the second-order equations

$$\ddot{r}_1 = b\dot{r}_2 e^{r_2} \quad (28)$$

and $\quad \ddot{r}_2 = B\dot{r}_1 e^{r_1} \ . \quad (29)$

From (26), (27) and (28) we can eliminate $r_2$ and $\dot{r}_2$ to write

$$\ddot{r}_1 = -(A + Be^{r_1})(a - \dot{r}_1). \quad (30)$$

A similar result for the second-order differential equation for $r_2$ reads

$$\ddot{r}_2 = -(a + be^{r_2})(A - \dot{r}_2). \quad (31)$$

For both (30) and (31) we can write equations similar to that in (11) and thus find their first integrals. In particular, from (30) we have

$$c(x, \dot{x}) = \dot{x} - Ax - Be^x + a\ln(\dot{x} - a). \quad (32)$$

Here $x$ and $\dot{x}$ stand for $r_1$ and $\dot{r}_1$. The choice $C(x, \dot{x}) = c(x, \dot{x})$ when used in (14) leads to the Lagrangian

$$L_1 = Ax + Be^x + (\dot{x} - a)\ln(\dot{x} - a). \quad (33)$$

One can verify that the Lagrangian $L_1$ in (33) via the Euler-Lagrangan equation gives (30). Similarly, the Lagrangian for (31) is given by

$$L_2 = ax + be^x + (\dot{x} - A)\ln(\dot{x} - A), \ x = r_2 \text{ and } \dot{x} = \dot{r}_2. \quad (34)$$

The expressions for $L_1$ and $L_2$ minus some gauge functions agree with those given in ref.17. A set of alternative nonstandard Lagrangians corresponding to the results in (33) and (34) can, in principle, be derived by choosing $C(x, \dot{x})$'s in terms of $c(x, \dot{x})$ in different ways. For example, if we take $C(x, \dot{x}) = c^2(x, \dot{x})$ we immediately obtain

$$L_1^a = \frac{1}{2}\dot{x}^2 - \frac{B^2}{2}e^{2x} + aBe^x \ln(\dot{x}-a) - \frac{a^2}{2}\ln(x-a)^2 - ABxe^x + aAx\ln(\dot{x}-a) - \frac{A^2}{2}x^2$$
$$- B\dot{x}e^x \ln(\dot{x}-a) + \frac{a}{2}\dot{x}\ln(\dot{x}-a)^2 - Ax\dot{x}\ln(\dot{x}-a) \tag{35}$$

and

$$L_2^a = \frac{1}{2}\dot{x}^2 - \frac{b^2}{2}e^{2x} + Abe^x \ln(\dot{x}-A) - \frac{A^2}{2}\ln(x-A)^2 - abxe^x + aAx\ln(\dot{x}-A) - \frac{a^2}{2}x^2$$
$$- b\dot{x}e^x \ln(\dot{x}-A) + \frac{A}{2}\dot{x}\ln(\dot{x}-A)^2 - ax\dot{x}\ln(\dot{x}-A). \tag{36}$$

Here the superscript $a$ on $L$ has been used merely to indicate that (35) and (36) stand for the alternative Lagrangians of those in (33) and (34) respectively. These Lagrangians belong to the standard class with velocity dependent potentials as appear in one very important type of force field, namely, the electromagnetic forces on moving charges [8].

**5. Concluding Remarks**

We considered the modified Emden-type equations obtained the expressions for the nonstandard Lagrangians by solving the associated inverse problem of the calculus of variations. In this context we also provided a family of inequivalent Lagrangians The modified Emden-type equation with friction quadratic in velocity and Lotka-Volterra model was found to follow from both standard and non-standard Lagrangians. See, for example, the results in (20), (35) and (36). That a physical system can have both standard and non-standard Lagrangian representation can also be demonstrated in an elementary level. Consider the equation of motion of a damped harmonic oscillator (DHO) written as
$$\ddot{x} + \gamma\dot{x} + x = 0. \tag{37}$$
It is well known the time-dependent standard Lagrangian for (37) is given by [20]
$$L = \frac{1}{2}e^{\lambda t}(\dot{x}^2 - x^2). \tag{38}$$
However, writing (37) as a system of two first-order differential equation one can obtain the first integrals of the equation as
$$c(x,\dot{x}) = \frac{1}{2}\ln(x^2 + \gamma x\dot{x} + \dot{x}^2) - \frac{\gamma}{\sqrt{4-\gamma^2}}\tan^{-1}\phi, \quad \phi = \frac{\gamma x + 2\dot{x}}{x\sqrt{4-\gamma^2}}, \gamma \neq 2 \tag{39}$$

and $\quad c(x,\dot{x}) = \ln(x+\dot{x}) + \dfrac{x}{x+\dot{x}}, \quad \gamma = 2 \tag{40}$

The Lagrangians corresponding to the constants of the motions in (39) and (40) are given by
$$L = \frac{\gamma \tan^{-1}\phi}{\sqrt{4-\gamma^2}} - \frac{1}{2}\ln(x^2 + \gamma x\dot{x} + \dot{x}^2) + 2\frac{\dot{x}}{x}\frac{\tan^{-1}\phi}{\sqrt{4-\gamma^2}} \tag{42}$$
and $\quad L = \ln(x+\dot{x}) \tag{43}$

respectively. From (38), (42) and (43) we confirm that the DHO can have both standard and non-standard Lagrandian representations. Moreover, it is interesting to note that, as with the modified Emden-type equation in (5), the DHO for $\gamma = 2$ also follows from a logarithmic Lagrangian. We conclude by noting that the model used by us provides an uncomplicated method to construct non-standard Lagrangian representations of mechanical systems from first principles.

**References**
[1] Santilli R M 1978 *Foundations of theoretical mechanics,* **I**, First Edition (Springer Verlag, New York)

[2] Cariñena J F, Rañada M F and Santander F 2005 J. Math. Phys. **46** 062703(17)
[3] Helmholtz H 1887 J Reine Angew. Math., **100** 137
[4] Olver P J *Applications of Lie Groups to Differential Equations* 1993 (Springer-Verlag, New York)
[5] Nucci M C and Leach P G L 2007 J. Math. Phys. **48** 123510
[6] Whittaker E T 1952 *A treatise on the analytical dynamics of particles and rigid bodies,* Third Edition, (Cambridge University Press, Cambridge, U. K.)
[7] Lopez, G 1996 Ann. Phys. (NY) **251,** 363
[8] Goldstein H 1998 *Classical Mechanics* 7th reprint (Narosa Pub. House, New Delhi, India)
[9]. Ince E L 1956 *Ordinary Differential equations* (Dover Pub., New York)
[10] Chandrasekher S 1957 *An Introduction to the study of Stellar Structure* (Dover Pub., New York)
[11] Musielak Z E 2008 J. Phys. A: Math. Theor. **41** 055305 (17)
[12] Chandrasekar V K, Santhilvelan M and Lakshmanan M 2005 Proc. R. Soc. London Ser. A **461** 2451
[13] Aparna Saha and Benoy Talukdar 2013 *On the non-standard Lagrangian equations* arXiv: 1301:2667
[14] Morandi G, Ferrario C, Vecchio G L, Marmo G and Rubano C 1990 Phys. Rep. **188** 147
[15] Ghosh S, Talukdar B, Das U and Saha A 2012 J. Phys. A: Math and Theor**. 45** 155207
[16] Trubatch S L and Franco A 1974 J. Theor. Biol**. 48** 299
[17] Nucci M C and Tamizhmani K M 2012 J. Nonlinear Math. Phys**. 19** 1250021
[18] Chandrasekar V K, Senthilvelan M and Lakshmanan M 2007 J. Phys. A **40,** 4717
[19] Reyes M A and H.C. Rosu H C 2008 J. Phys. A: Math. Theor. **41** 28520
[20] Caldirola P 1941 Nuovo. Cim. **18** 393 ; Kanai E 1948 Prog. Theor. Phys. **20** 440